\documentclass[12pt]{article}

\renewcommand{\d}{{\rm d}}

\newcommand{\ve}{\varepsilon}

\begin{document}
\title{Faddeev gravity action on the piecewise constant fundamental vector fields}
\author{V.M. Khatsymovsky \\
 {\em Budker Institute of Nuclear Physics} \\ {\em of Siberian Branch Russian Academy of Sciences} \\ {\em
 Novosibirsk,
 630090,
 Russia}
\\ {\em E-mail address: khatsym@inp.nsk.su}}
\date{}
\maketitle
\begin{abstract}
In the Faddeev formulation of gravity, the metric is regarded as composite field, bilinear of $d = 10$ 4-vector fields.
We derive the minisuperspace (discrete) Faddeev action by evaluating the Faddeev action on the spacetime composed of the (flat) 4-simplices with constant 4-vector fields. This is an analog of the Regge action obtained by evaluating the Hilbert-Einstein action on the spacetime composed of the flat 4-simplices. One of the new features of this formulation is that the simplices are not required to coincide on their common faces. Also an analog of the Barbero-Immirzi parameter $\gamma$ can be introduced in this formalism.
\end{abstract}

keywords: Einstein gravity; Regge calculus; composite metric; Faddeev gravity; discrete gravity

\section{Introduction}
Regge calculus \cite{Reg} is a minisuperspace formulation of gravity, that is, exact general relativity (GR) on a family of the metric fields (piecewise flat ones) sufficiently large to approximate any metric with any accuracy. Spacetime is taken as the set of the flat 4d tetrahedra. Let $\sigma^n$ be $n$-dimensional tetrahedron or $n$-simplex. Then the Einstein action is a sum over triangles $\sigma^2$,
\begin{equation}
\frac{1}{2} \int R \sqrt{g} {\rm d}^4 x = \sum_{\sigma^2}A_{\sigma^2} \alpha_{\sigma^2}.
\end{equation}

\noindent Here, $A_{\sigma^2}$ is area of the triangle $\sigma^2$, $\alpha_{\sigma^2}$ is the defect angle on the triangle $\sigma^2$. Regge calculus is expected to be the baze for quantizing GR \cite{RegWil,Ham,cdt} difficult to implement by the other methods because of the formal nonrenormalizability of GR.

Let us consider some another minisuperspace theory based on the Faddeev formulation of gravity \cite{Fad}. In the Faddeev gravity, the metric is composed of $d = 10$ 4-vector fields,
\begin{equation}
g_{\lambda\mu} = f^A_\lambda f_{\mu A}.
\end{equation}

\noindent Here, $\lambda, \mu$, \dots = 1, 2, 3, 4; $ ~~~ A, B$, \dots = 1, \dots, $d$. Simple example: locally, 4d Riemannian space can be considered as a hyper-surface in the 10d Euclidean space. If $f^A (x)$ were its coordinates, then we would have
\begin{equation}
f^A_\lambda = \partial f^A / \partial x^\lambda.
\end{equation}

In the Faddeev gravity, however, $f^A_\lambda (x)$ are independent variables. We can introduce an alternative connection
\begin{equation}
\tilde{\Gamma}_{\lambda \mu\nu} = f^A_\lambda f_{\mu A, \nu} ~ (f_{\mu A, \nu} \equiv \partial_\nu f_{\mu A}), ~ \tilde{\Gamma}^\lambda_{\mu\nu} = g^{\lambda\rho} \tilde{\Gamma}_{\rho \mu \nu},
\end{equation}

\noindent its curvature is
\begin{eqnarray}
K^\lambda_{\mu \nu \rho} = \tilde{\Gamma}^\lambda_{\mu\rho, \nu} - \tilde{\Gamma}^\lambda_{\mu\nu, \rho} + \tilde{\Gamma}^\lambda_{\sigma\nu} \tilde{\Gamma}^\sigma_{\mu\rho} - \tilde{\Gamma}^\lambda_{\sigma\rho} \tilde{\Gamma}^\sigma_{\mu\nu} \nonumber \\ = \Pi^{AB} (f^\lambda_{A, \nu} f_{\mu B, \rho} - f^\lambda_{A, \rho} f_{\mu B, \nu}).
\end{eqnarray}

\noindent Here,
\begin{equation}
\Pi_{AB} = \delta_{AB} - f^\lambda_A f_{\lambda B}
\end{equation}

\noindent is a projector. Note that it makes usual and covariant derivatives equivalent,
\begin{equation}
\Pi^{AB} f_{\lambda B, \mu} = \Pi^{AB} \nabla_\mu f_{\lambda B}.
\end{equation}

Then we can write out the Faddeev action,
\begin{eqnarray}\label{Faddeev}
S = \int g^{\lambda \nu} g^{\mu \rho} K_{\lambda \mu \nu \rho} \sqrt {g} \d^4 x \nonumber \\ + \frac{1}{2 \gamma} \int \epsilon^{\lambda \mu \nu \rho} K_{\lambda \mu \nu \rho} \d^4 x \nonumber \\
= \int \Pi^{AB} \left [ (f^\lambda_{A, \lambda} f^\mu_{B, \mu} - f^\lambda_{A, \mu} f^\mu_{B, \lambda}) \sqrt {g} \right. \nonumber \\ \left. - \frac{1}{\gamma} \epsilon^{\lambda \mu \nu \rho} f_{\lambda A, \mu} f_{\nu B, \rho} \right ] \d^4 x.
\end{eqnarray}

\noindent Here we have generalized it by adding the P-odd $1/\gamma$-term where $\gamma$ is an analog \cite{Kha0} of the Barbero-Immirzi parameter \cite{Barb}-\cite{Fat} in the usual Einstein gravity (in the connection representation).

The variation of the action reads
\begin{eqnarray}
\frac{\delta S}{2 \sqrt{g} \delta f^\lambda_A} = f^A_\mu \left ( K^\mu_\lambda - \frac{1}{2} \delta^\mu_\lambda K - \frac{\epsilon^{\mu \nu \rho \sigma}}{2 \gamma \sqrt{g}} K_{\lambda \nu \rho \sigma} \right ) \nonumber \\
+ \Pi^{AB} \left [ f^\nu_{B, \nu} T^\mu_{\lambda \mu} + f^\nu_{B, \mu} T^\mu_{\nu \lambda} + f^\nu_{B, \lambda} T^\mu_{\mu \nu} \right. \nonumber \\ \left. + \frac{\epsilon^{\tau \mu \nu \rho}}{2 \gamma \sqrt{g}} (g_{\lambda \sigma} g_{\kappa \tau} - g_{\lambda \tau} g_{\kappa \sigma}) f^{\kappa}_{B, \rho} T^\sigma_{\mu \nu} \right ].
\end{eqnarray}

\noindent Here, the $\Pi^{AB}$-part of the field eqs $\delta S = 0$ turns out to be a linear homogeneous system for torsion $T^\lambda_{ \mu \nu } = f^\lambda_A (f^A_{\mu, \nu} - f^A_{\nu, \mu})$ with a nondegenerate $24 \times 24$ matrix leading to $T^\lambda_{\mu \nu} = 0$. This gives $\tilde{\Gamma}^\lambda_{\mu\nu} = \Gamma^\lambda_{\mu\nu},$ the Cristoffel. Then $K^\lambda_{\mu \nu \rho} = R^\lambda_{\mu \nu \rho},$ the Riemann for which $\epsilon^{\mu \nu \rho \sigma} R_{\lambda \nu \rho \sigma} \equiv 0,$ and the $f^A_\mu$-part of the field eqs turns out to be the Einstein eqs.

\section{The piecewise constant fields}\label{sec2}

In GR, taking the metric as
\begin{equation}
\d s^2 = g_{nn} (\d x^n)^2 + g_{\alpha \beta} \d x^\alpha \d x^\beta
\end{equation}

\noindent where $n$ is any of 1, 2, 3, 4, and $\alpha, \beta, \gamma, ... \neq n$, we can separate out the derivatives over $x^n$ squared in the action as follows,
\begin{eqnarray}
\int R \sqrt{g} \d^4 x = \hspace{50mm} \nonumber \\ \frac{1}{4 } \int (g^{\alpha\gamma}g^{\beta\delta} - g^{\alpha\beta}g^{\gamma\delta})g_{\alpha\beta , n}g_{\gamma\delta , n}g^{nn}\sqrt{g}\d^4x + ....
\end{eqnarray}

\noindent If the metric is taken to be piecewise constant (with $g_{\lambda \mu} = const$ inside the 4-simplices),  this implies continuity condition for the induced on the faces metric,
\begin{equation}
g_{\alpha \beta}|_{x^n = x^n_0 + 0} - g_{\alpha \beta}|_{x^n = x^n_0 - 0} = 0 ~~~ \forall ~ x^n_0,
\end{equation}

\noindent if some 3-face is described locally by $x^n = x^n_0$. That is, the simplices should coincide on their common faces.

In the Faddeev gravity, the fields $f^A_\lambda$ can be taken to be piecewise constant. The action
\begin{eqnarray}
S = \int \Pi^{AB} \left [ (f^\lambda_{A, \lambda} f^\mu_{B, \mu} - f^\lambda_{A, \mu} f^\mu_{B, \lambda}) \sqrt {g} \right. \nonumber \\ \left. - \frac{1}{\gamma} \epsilon^{\lambda \mu \nu \rho} f_{\lambda A, \mu} f_{\nu B, \rho} \right ] \d^4 x
\end{eqnarray}

\noindent does not contain the square of any derivative and does not require continuity condition on $f^A_\lambda (x )$ and thus on the metric $g_{\lambda\mu} = f^A_\lambda f_{\mu A}$. The simplices are not required to coincide on their common faces, and we can take $f^A_\lambda (x ) = const$ independently in these 4-simplices.

\section{The discrete Faddeev action}

To evaluate the discrete action \cite{Kha}, we divide ${\rm I \hspace{-2pt} R}^4$ as the set of points $x = (x^1, x^2, x^3, x^4)$ by the hypersurfaces $a_\lambda x^\lambda + b = 0$ (mathematical hyperplanes) into polytopes, in particular, 4-simplices. $f^A_\lambda$ are independent in these 4-simplices.

Then we note that the expression
\begin{equation}
Q_{AB} = f^\lambda_{A, \lambda} f^\mu_{B, \mu} - f^\mu_{A, \lambda} f^\lambda_{B, \mu}
\end{equation}

\noindent 1) is zero if $f^\lambda_A$ depends only on one coordinate: contribution of the 3-simplices vanishes, and $Q_{AB} = const \cdot \delta (x^1) \delta (x^2)$ in a neighborhood of the 2-simplex $\sigma^2$ at $x^1 = 0, x^2 = 0$ (triangle in $x^3, x^4$-plane); \\ 2) is a full derivative
\begin{equation}
Q_{AB} = \partial_\lambda (f^\lambda_A \partial_\mu f^\mu_B - f^\mu_A \partial_\mu f^\lambda_B )
\end{equation}

\noindent so that
\begin{equation}
\int Q_{AB} \d x^1 \d x^2 = \oint_C (f^1_A \d f^2_B - f^2_A \d  f^1_B)
\end{equation}

\noindent where the contour $C$ encircles the triangle $\sigma^2$. This gives
\begin{eqnarray}
2 Q_{AB} = \delta (x^1 ) \delta (x^2 ) \sum^n_{i=1} [ f^1_A ( \sigma^4_i ) f^2_B ( \sigma^4_{i+1} ) \nonumber \\ - f^1_A ( \sigma^4_{i+1} ) f^2_B ( \sigma^4_i ) ] + (A \leftrightarrow B)
\end{eqnarray}

\noindent where the 4-simplices $\sigma^4_i, i = 1 \div n$ surround the triangle $\sigma^2$. The action is
\begin{eqnarray}
\frac{1}{2} \Pi^{AB} (\sigma^2 ) \sum^n_{i=1} \left \{ \left [ f^1_A ( \sigma^4_i ) f^2_B ( \sigma^4_{i+1} ) \right. \phantom{\frac{1}{\gamma}} \right. \nonumber \\ \left. - f^1_A ( \sigma^4_{i+1} ) f^2_B ( \sigma^4_i ) \right ] \sqrt{g (\sigma^2 )} \nonumber \\
\left. - \frac{1}{\gamma} \left [ f_{4 A }( \sigma^4_i ) f_{3 B} ( \sigma^4_{i+1} ) - f_{4 A } ( \sigma^4_{i+1} ) f_{3 B} ( \sigma^4_i ) \right ] \right \} \nonumber \\ \cdot ( \Delta x^3_{\sigma^1_3} \Delta x^4_{\sigma^1_4} - \Delta x^3_{\sigma^1_4} \Delta x^4_{\sigma^1_3} )
\end{eqnarray}

\noindent where $\sigma^2$ is formed by the edges $\sigma^1_3, \sigma^1_4$ with the 4-vectors $\Delta x^\lambda_{\sigma^1_3}$, $\Delta x^\lambda_{\sigma^1_4}$.

\section{An invariant form of the action}

We can rewrite this action in some invariant variables. This invariance is that one w. r. t. the choice of the coordinates $x^\lambda$ assigned to the vertices. Choose some two edges $\sigma^1_1, \sigma^1_2$ in addition to the above $\sigma^1_3, \sigma^1_4$ spanning $\sigma^2$ so that $\sigma^1_1, \sigma^1_2, \sigma^1_3, \sigma^1_4$ span some $\sigma^4_{i_0} \supset \sigma^2$. The difference between $x^\lambda$ of the endpoints
of $\sigma^1_\mu$ is $\Delta x^\lambda_{\sigma^1_\mu}$. Invariant edge variables $f^{\sigma^1_\lambda}_A$ or $f_{\sigma^1_\lambda}^A$ are defined by
\begin{equation}
f^\lambda_A = \sum_\mu f^{\sigma^1_\mu}_A \Delta x^\lambda_{\sigma^1_\mu}, ~~~ f^A_{\sigma^1_\lambda } = f^A_\mu \Delta x^\mu_{\sigma^1_\lambda }.
\end{equation}

\noindent Then we can write the contribution of $\sigma^2$ to the action,
\begin{eqnarray}
\frac{1}{2} \Pi^{AB} (\sigma^2 ) \sum^n_{i=1} \left \{ \left [ f^{\sigma^1_1}_A ( \sigma^4_i ) f^{\sigma^1_2}_B ( \sigma^4_{i+1} ) \right. \phantom{\frac{1}{\gamma}} \right. \nonumber \\ \left. - f^{\sigma^1_1}_A ( \sigma^4_{i+1} ) f^{\sigma^1_2}_B ( \sigma^4_i ) \right ] \sqrt{\det \| g_{\sigma^1_\lambda \sigma^1_\mu} \|} \nonumber \\
- \frac{1}{\gamma} \left [ f_{\sigma^1_4 A }( \sigma^4_i ) f_{\sigma^1_3 B} ( \sigma^4_{i+1} ) \right. \nonumber \\ \left. \phantom{\frac{1}{\gamma}} \left. - f_{\sigma^1_4 A } ( \sigma^4_{i+1} ) f_{\sigma^1_3 B} ( \sigma^4_i ) \right ] \right \}
\end{eqnarray}

\noindent where $\Pi^A_B \! = \! \delta^A_B - \sum_\lambda f^A_{\sigma^1_\lambda } f_B^{\sigma^1_\lambda }$, $ g_{\sigma^1_\lambda \sigma^1_\mu} \! = \! f^A_{\sigma^1_\lambda } f_{ \sigma^1_\mu A }$ (metric edge components).

\section{Restoring the continuum action from the discrete minisuperspace one}

We can obtain the continuum Faddeev action from the found discrete minisuperspace one. This reverse transition is intended to show that the information encoded in the minisuperspace formulation is sufficient to reproduce the essential degrees of freedom of the continuum theory. This procedure is analogous to the work \cite{Fein} for the usual GR where the Regge action has been shown to tend to the Hilbert-Einstein one if simplicial decomposition of the given smooth manifold is made finer and finer. To do this, we choose the fields to approximate some fixed smooth $f^A_\lambda (x )$ arbitrarily closely by making decomposition of spacetime into the 4-simplices finer and finer (the coordinate steps along the edges $\Delta x^\lambda_{\sigma^1} \to 0$ ). Then we choose the discrete variable $f^\lambda_A (\sigma^4 )$ to be $f^\lambda_A (x )$ at $x = x_{\sigma^4}$, some central point in $\sigma^4$. The procedure is most easily illustrated by the case $n=4$:
\begin{eqnarray}
\sum^4_{i=1} \left [ f^1_A ( \sigma^4_i ) f^2_B ( \sigma^4_{i+1} ) - f^1_A ( \sigma^4_{i+1} ) f^2_B ( \sigma^4_i ) \right ] \nonumber \\ = \Delta_u f^1_A \Delta_v f^2_B - \Delta_v f^1_A \Delta_u f^2_B
\end{eqnarray}

\noindent where $\Delta_u f \equiv f(\sigma^4_3 ) - f(\sigma^4_1 )$ and $\Delta_v f \equiv f(\sigma^4_4 ) - f(\sigma^4_2 )$ in the continuum limit become derivatives. This gives us the basic structure (some two-dimensional Jacobian) which being summed over the different contravariant components $f^\lambda_{\dots}$ and variations between the different pairs of simplices $\Delta f = f (\sigma^4 ) - f (\sigma^{4 \prime } )$ just leads to $f^\lambda_{A, \lambda} f^\mu_{B, \mu} - f^\mu_{A, \lambda} f^\lambda_{B, \mu}$ and thus to the Faddeev action. The $1 / \gamma$-part of the action is considered completely analogously.

\section{Possibility of the cubic decomposition of spacetime}

A new interesting property of the Faddeev formulation as compared to the Regge calculus is the possibility to decompose the curved spacetime into the (flat) cubes. Consider decomposition of ${\rm I \hspace{-2pt} R}^4$ into the rectangular parallelepipeds or cuboids
\begin{eqnarray}
\{ x | x^\lambda_0 + n^\lambda \ve^\lambda \leq x^\lambda \phantom{\leq x^\lambda_0 + (n^\lambda + 1) \ve^\lambda, \lambda = 1, 2, 3} \nonumber \\ \phantom{n^\lambda \ve^\lambda \leq x^\lambda} \leq x^\lambda_0 + (n^\lambda + 1) \ve^\lambda, \lambda = 1, 2, 3, 4 \}.
\end{eqnarray}

\noindent Then $f^\lambda_A (x )$ is assumed to be constant in each of these cuboids. Above derivation of the minisuperspace Faddeev action remains the same if $\sigma^4$ are not 4-simplices but cuboids. In Regge calculus, the continuity of metric on the faces leads to that the spacetime composed of flat cuboids is flat as well. In the Faddeev gravity, continuity of $f^\lambda_A (x )$ is not required, and the flat cuboids can be used for modeling the curved spacetime. The cuboid action looks just as the naively discretized action or the continuum one (\ref{Faddeev}) in which the derivatives are substituted by their finite difference counterparts. It is much simpler than the simplicial one and, at the same time, it is a minisuperspace action.

\section{Conclusion}

To summarize, we have constructed some discrete formalism analogous to the Regge calculus, but based not on the usual GR, but on its Faddeev form. Some new features of this formalism are the following. \\
1) The invariant physical variables are edge 10-vectors $f^A_{\sigma^1} (\sigma^4 )$ (or $f^A_{\sigma^1} (\sigma^4 )$), independent for the different 4-simplices $\sigma^4$ containing the edge $\sigma^1$. \\
2) The 4-simplices do not necessarily coincide on their common faces. \\
3) $1/\gamma$-term is present which is an analogue of the so called Barbero-Immirzi parity odd term in the connection representation of the GR action.

An advantage of the discrete Faddeev formalism is the possibility to use cubic decomposition instead of the simplicial one for modeling the curved spacetime due to possibility of the field $f^\lambda_A$ discontinuities in the Faddeev gravity. The cuboid action looks simple as the naive discretization of the continuum one.

Also this possibility of the field $f^\lambda_A$ discontinuities allows one to evaluate the area spectrum of any surface as the sum of spectra of independent triangles. In blackhole physics, reasonable physical arguments require a discrete area spectrum, but up to now, only one gravity theory (LQG) allows to obtain it \cite{Ash} - \cite{ABCK}. We hope that the discrete Faddeev gravity approach is also able to handle this task.

The present work was supported by the Ministry of Education and Science of the Russian Federation.


\begin{thebibliography}{99}
\bibitem{Reg}
 Regge T. 1961 {\it Nuovo Cimento} {\bf 19} 558
\bibitem{RegWil}
 Regge T. and Williams R. M. 2000 {\it Journ. Math. Phys.} {\bf 41} 3964  [arXiv:0012035 [gr-qc]].
\bibitem{Ham}
 Hamber H. W. 2009 {\it Gen. Rel. Grav.} {\bf 41} 817  [arXiv:0901.0964 [gr-qc]].
\bibitem{cdt}
 Ambjorn J., Goerlich A., Jurkiewicz J., and Loll R. 2012 {\it Physics Reports} {\bf 519} 127  [arXiv:1203.3591 [hep-th]].
\bibitem{Fad}
 Faddeev L. D. 2011 {\it Theor. Math. Phys.} {\bf 166} 279
\bibitem{Kha0}
 Khatsymovsky V. M. 2013 {\it Class.Quant.Grav.} {\bf 30} 095006  [arXiv:1201.0806 [gr-qc]].
\bibitem{Barb}
 Barbero J. F. 1995 {\it Phys. Rev. D} {\bf 51} 5507  [arXiv:9410014 [gr-qc]].
\bibitem{Holst}
 Holst S. 1996
 {\it Phys. Rev. D} {\bf 53} 5966  [arXiv:9511026 [gr-qc]].
\bibitem{Imm}
 Immirzi G. 1997 {\it Nucl. Phys. Proc. Suppl.} {\bf 57} 65  [arXiv:9701052 [gr-qc]].
\bibitem{Fat}
 Fatibene L., Francaviglia M., and Rovelli C. 2007 {\it Class. Quantum Grav.} {\bf 24} 4207  [arXiv:0706.1899 [gr-qc]].
\bibitem{Kha} Khatsymovsky V. M. 2014 {\it Mod. Phys. Lett.} A {\bf 29} 1450141  [arXiv:1408.6375 [gr-qc]].
\bibitem{Fein}
 Feinberg G., Friedberg R., Lee T. D., and Ren M. C. 1984 {\it Nucl. Phys.} B {\bf 245} 343
\bibitem{Ash}
 Ashtekar A., Rovelli C. and Smolin L. 1992 {\it Phys. Rev. Lett.} {\bf 69} 237  [arXiv:9203079 [hep-th]].
\bibitem{Loll}
 Loll R. 1997 {\it Class. Quant. Grav.} {\bf 14} 1725  [arXiv:9612068 [gr-qc]].
\bibitem{ABCK}
 Ashtekar A., Baez J., Corichi A., and Krasnov K. 1998 {\it Phys.Rev.Lett.} {\bf 80} 904  [arXiv:9710007 [gr-qc]].
\end{thebibliography}
\end{document}